\documentclass[aps, prl,amsmath,amssymb,twocolumn,floatfix]{revtex4}

\newif\ifpdf\ifx\pdfoutput\undefined\pdffalse\else\pdfoutput=1\pdftrue\fi

\usepackage{graphicx}
\usepackage{bm}
\usepackage{epsfig}

\newcommand{\p}{^{0}}
\newcommand{\cl}{_{\rm cl}}
\renewcommand{\sp}{_{\rm sp}}
\newcommand{\DP}{\Delta P}

\begin{document}

\title{\bf Finite-size scaling and particle size cutoff effects in
phase separating polydisperse fluids}

\author{Nigel B. Wilding}
\affiliation{Department of Physics, University of Bath, Bath BA2 7AY, United Kingdom.}

\author{Peter Sollich}
\author{Moreno Fasolo}
\affiliation{King's College London, Department of Mathematics, Strand,
London WC2R 2LS, United Kingdom.}

\date{\today}

\begin{abstract} 

We study the liquid-vapor phase behaviour of a polydisperse fluid using
grand canonical simulations and moment free energy calculations. The
strongly nonlinear variation of the fractional volume of liquid across
the coexistence region prevents naive extrapolation to detect the cloud
point. We describe a finite-size scaling method which nevertheless
permits accurate 
determination of cloud points and spinodals from simulations of a
single system size. By varying a particle size cutoff we find that the
cloud point density is highly sensitive to the presence of rare large
particles; this could affect the reproducibility of experimentally
measured phase behavior in colloids and polymers. 


\end{abstract} 
\maketitle
\setcounter{totalnumber}{10}

Many complex fluids such as colloidal dispersions, liquid crystals and
polymer solutions are inherently polydisperse in character: their
constituent particles have an essentially continuous range of size,
shape or charge. Polydispersity is of significant practical importance
because it can affect material properties in applications ranging from
coating technologies and foodstuffs to polymer processing
\cite{FREDRICKSON}. However, our understanding of the fundamental
properties of polydisperse fluids remains very limited compared with
what we know about their monodisperse counterparts. The challenge
arises because a polydisperse fluid is effectively a mixture of an {\em
infinite} number of particle species. Labelling each by the value of
its polydisperse attribute $\sigma$, the state of the system (or any of
its phases) has to be described by a density {\em distribution}
$\rho(\sigma)$, with $\rho(\sigma)d\sigma$ the number density of
particles in the range $\sigma\ldots \sigma+d\sigma$. The most common
experimental situation is that in which the form of the overall or
``parent'' distribution $\rho\p(\sigma)$ is fixed by the synthesis of
the fluid, and only its scale can vary depending on the proportion of
the sample volume occupied by solvent. One can then write
$\rho\p(\sigma)=n\p f\p(\sigma)$ where $f\p(\sigma)$ is the normalized
parent shape function and $n\p=N/V$ the overall particle number
density. Varying $n\p$ at a given temperature corresponds to scanning a
``dilution line'' of the system. 

A central issue in the physics of polydisperse fluids is the nature of
their phase behaviour: in order to process a polydisperse fluid one
needs to know under which conditions it will demix and what phases will
result. However, the phase behaviour of polydisperse systems can be
considerably richer than that of monodisperse systems, due to the
occurrence of {\em fractionation} \cite{EVANS,GHOSH,ERNE05}: at
coexistence, particles of each $\sigma$ may partition themselves
unevenly between two or more ``daughter'' phases as long as--due to
particle conservation--the overall density distribution
$\rho\p(\sigma)$ of the parent phase is maintained. As a
consequence, the conventional fluid-fluid binodal of a monodisperse
system splits into a cloud curve marking the onset of coexistence, and
a shadow curve giving the density of the incipient phase; the critical
point appears at the intersection of these curves rather than at the
maximum of either~\cite{SOLLICH02}.

\begin{figure}[h]
\includegraphics[width=8.0cm,clip=true]{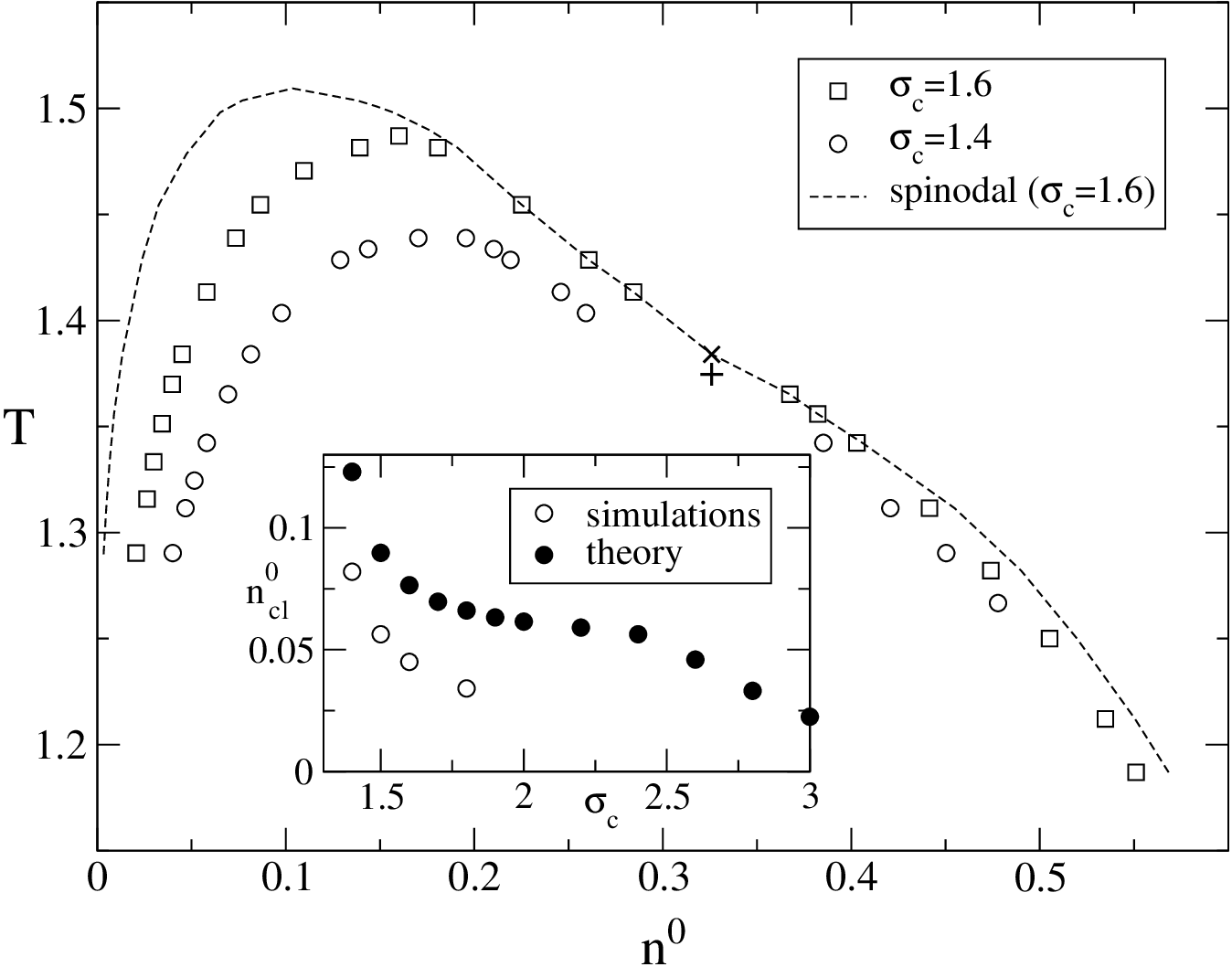}
\caption{Comparison of cloud curves for $\sigma_c=1.4$ and
$\sigma_c=1.6$. The critical points for $\sigma_c=1.6$ ($\times$) and
$\sigma_c=1.4$ ($+$) are marked. Also shown is the spinodal (limit of
metastability) for $\sigma_c=1.6$. The inset displays the variation of
the gas cloud point density $n\p\cl$ at $T=T_c$ as a function of
$\sigma_c$, as obtained from GCE simulations (open symbols) and MFE
theory (filled symbols).
\label{fig:cloudcurve}
}
\end{figure}

In this letter we describe a joint simulation and theoretical study of
a model polydisperse Lennard-Jones fluid in which the size of the
particles influences not only the length-scale but also the strength
$\epsilon_{ij}$ of the interparticle potentials, as defined
in~(\ref{uij}) below.  For the case of size-independent interaction
strengths, the critical point lies very close to the maximum of the
cloud curve \cite{WILDING04}, whereas for the present model we find
that it is substantially below (see Fig.~\ref{fig:cloudcurve}), as is
observed in many experiments on complex fluids (see
e.g.~\cite{DOBASHI}) and simplified theoretical
calculations~\cite{BAUS}. At the critical temperature, $T_c$, there
then exists a finite density range where phase separation occurs on the
dilution line. Most results shown below will be at
this temperature; note that we will be interested mainly in the
low-density part of the coexistence region rather than the critical
effects at the other end, using $T_c$ merely as a convenient
temperature scale.

The simulations were performed within the grand canonical ensemble
(GCE). This is particularly useful for polydisperse systems, where it
permits sampling of many different realizations of the particle size
distribution while catering naturally for fractionation effects.
Operationally, we ensure that the ensemble averaged density
distribution always equals the desired parent distribution
$\rho\p(\sigma)$ by controlling an imposed chemical potential
distribution $\mu(\sigma)$. A combination of novel and existing
techniques~\cite{WILDING03} are required to tune $\mu(\sigma)$ such as
to track the dilution line, i.e.\ to vary the parent density $n\p$ but
not its shape $f\p(\sigma)$.

\begin{figure}[h]
\includegraphics[width=8.5cm,clip=true]{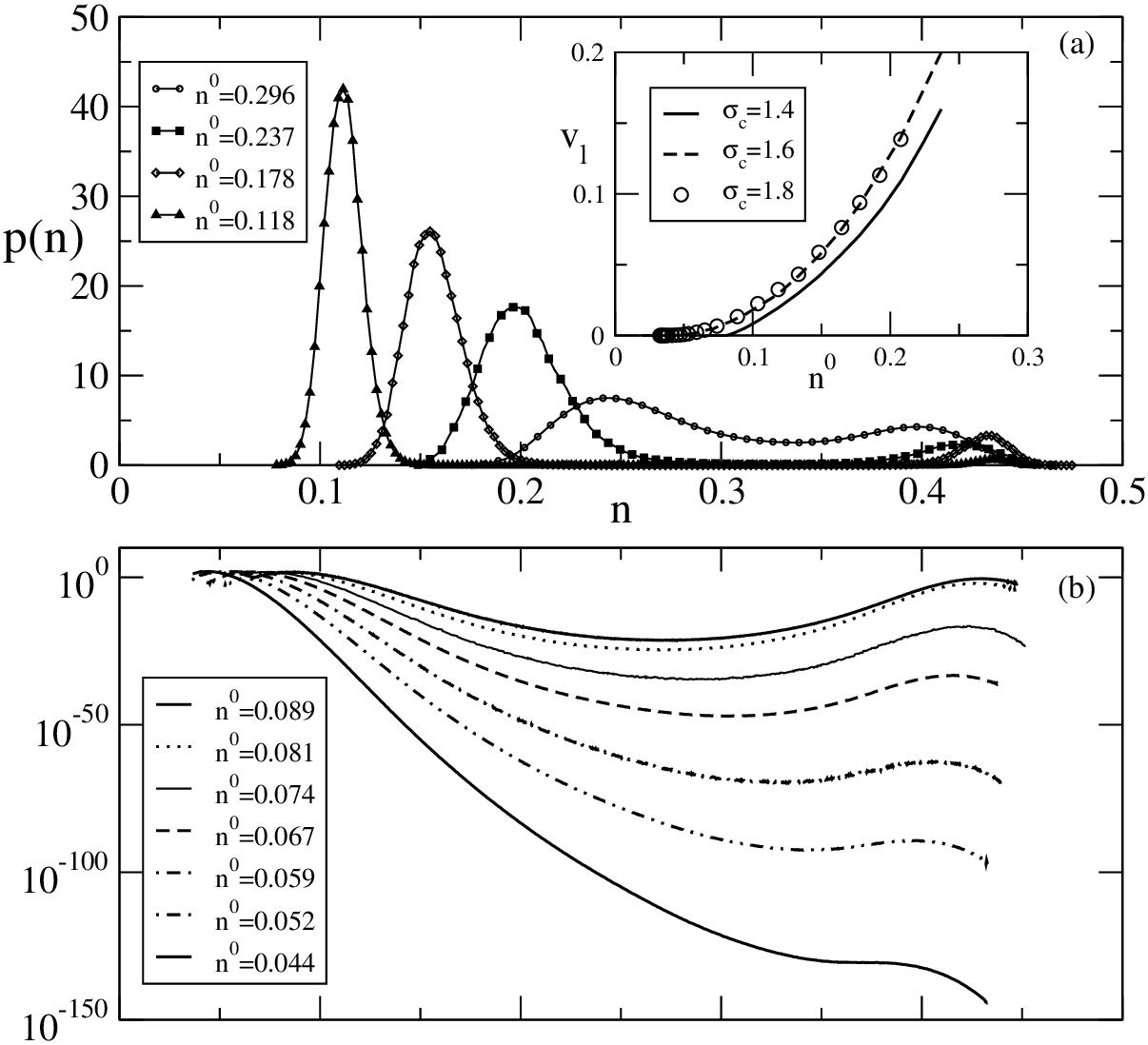}
\caption{Number density distribution $p(n)$ at $T=T_c$ for parent
densities $n\p$ as indicated and for particle size cutoff 
$\sigma_c=1.4$. The system size is $L=15\bar\sigma$. See text after
eq.~(\ref{uij})) for definitions of $\sigma_c$ and $\bar\sigma$.
(a) Linear and (b) log scale. Inset: Liquid fractional
volume $v_{\rm l}$ versus $n\p$, for $\sigma_c=1.4$, 1.6, 1.8.}
\label{fig:coexdists}
\end{figure}
In the GCE simulations, the number density $n$ is a fluctuating
variable with average equal to $n\p$. Its distribution $p(n)$, shown
in Fig.~\ref{fig:coexdists} for a range of parent densities $n\p$ at
$T=T_c$, is a key observable. In the coexistence region it has two
distinct peaks, which we sample using multicanonical
preweighting~\cite{BERG}. The weight under the low and high
density peaks corresponds respectively to the fractional volumes $v_{\rm g}$
and $v_{\rm l}$ that would be occupied by gas and liquid in the
corresponding canonical ensemble. As expected, the peaks separate and
the valley between them deepens as we move away from the critical
point by decreasing $n\p$. Concomitant with this is a gradual
transfer of weight from the liquid to the gas peak. Finally the liquid
peak disappears, at exponentially small values of $v_{\rm l}$ visible only
on a log scale (Fig.~\ref{fig:coexdists}(b)).

The observed variation of $p(n)$ raises the question of how to detect
the cloud point $n\p\cl$, defined as the lowest parent density $n\p$
where stable phase coexistence occurs. In a monodisperse system this is
straightforward because the cloud point also gives the density of the
gas phase, which remains constant throughout the coexistence region.
One then simply detects the point where the gas and liquid peaks have
the same weight, i.e.\ $r=v_{\rm l}/v_{\rm g}=1$, and measures the gas density
there. (The criterion $r=1$ has the added advantage of leading to only
exponentially small finite-size corrections to the value of $\mu$ at
coexistence~\cite{BORGS92}). However, this method fails in a
polydisperse system because fractionation causes the densities and size
distributions of the coexisting phases to vary with
$n\p$~\cite{SOLLICH02}. One could attempt to locate the cloud point
instead by extrapolating in $n\p$ to the point where $v_{\rm l}\to
0$~\cite{WILDING04}. But in our system the dependence of $v_{\rm l}$ on $n\p$
is so strongly nonlinear---another effect of fractionation, see inset
of Fig.~\ref{fig:coexdists}---that the resulting cloud point estimates
would have very large error bars. Indeed, on a linear plot of $v_{\rm l}$
versus $n\p$ as shown in Fig.~\ref{fig:coexdists}(a) 
the effects of the particle size cutoff $\sigma_c$ which
our more careful 
analysis will reveal (see Fig.~\ref{fig:cloudcurve} below) are
essentially invisible.

To make progress, we analyse the finite-size scaling of $p(n)$. As the
linear system size $L$ grows at fixed $n\p$ and $T$, the peaks in
$p(n)$ will narrow around the densities of gas and liquid,
respectively, and the size distributions averaged over configurations
from each peak will tend to those in the coexisting phases. The ratio
$r=v_{\rm l}/v_{\rm g}$ is determined by the difference in the grand potential;
this is directly related to the pressure $P$ so that $r=\exp(\beta L^d
\DP)$ for large $L$ where $\beta=1/k_{\rm B}T$ and $\DP = P_{\rm l}-P_{\rm g}$. The
criterion for stable coexistence at given fixed $n\p$ is that $r$ must
have a finite value as $L\to\infty$; the pressure difference then has
to scale as $\DP \sim L^{-d}$ except in the special case $r=1$ (see
above).

\begin{figure}[h]
\includegraphics[width=8.0cm,clip=true]{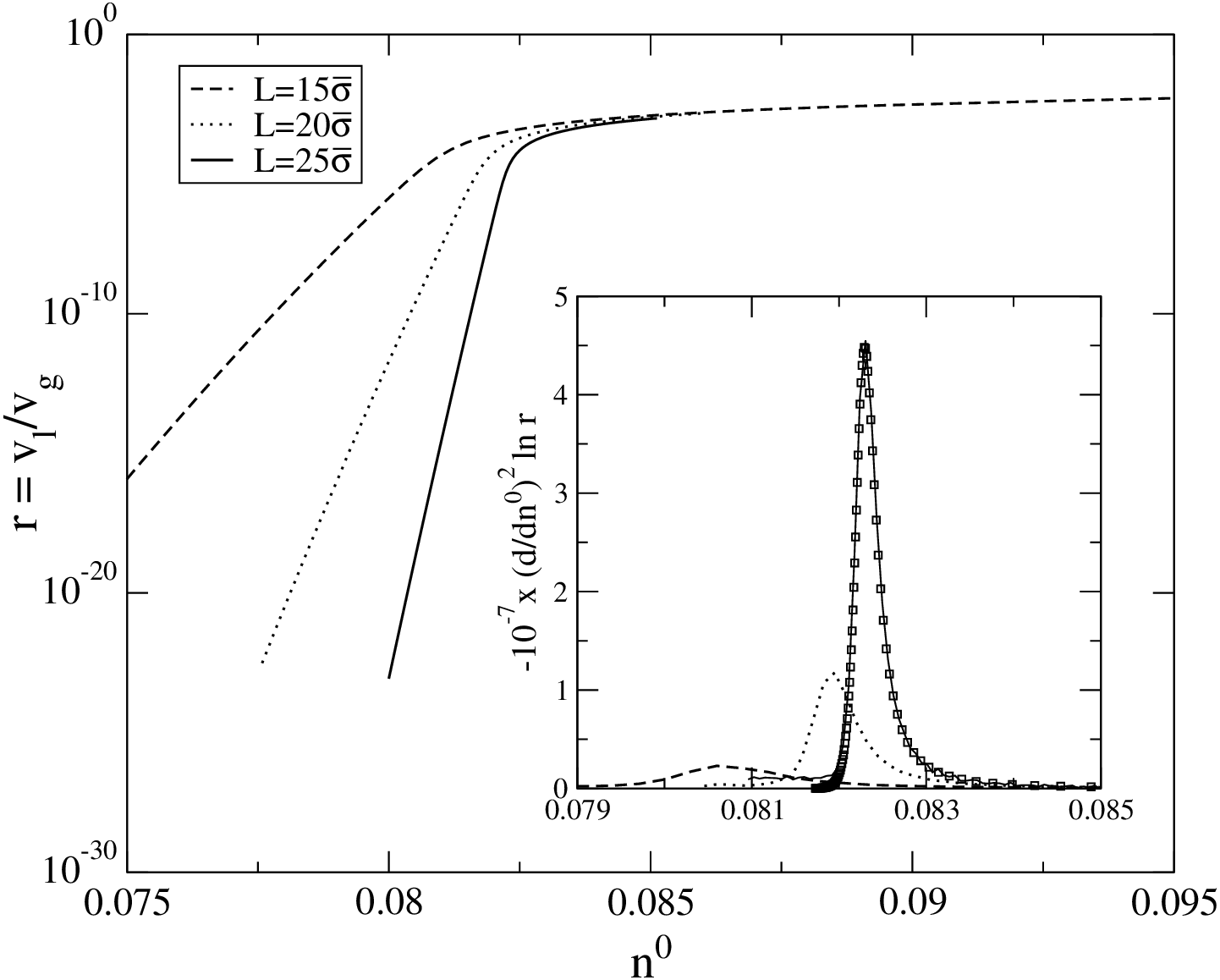}
\caption{Ratio $r$ of liquid to gas fractional volumes on approach to
the cloud point at $T=T_c$ for $\sigma_c=1.4$. The inset shows the
(negative, scaled) second derivative of $\ln r$ w.r.t. $n\p$. The peak
position gives an estimate 
of the cloud point density. Squares indicate the scaled master
curve~(\protect\ref{eq:master}).}
\label{fig:fss}
\end{figure}

For finite $L$, metastable coexistence can still be observed in the
density region $n\p<n\p\cl$ where $\Delta P=O(1)$, but here $r$ will be
exponentially small. Fig.~\ref{fig:fss} shows this effect clearly: $r$
is independent of $L$ for sufficiently large $n\p$, but the curves
depart rapidly from each other
(note the log-scale) for smaller $n\p$. The cloud
point separates the two regimes, permitting the estimate $n\p\cl\approx
0.0825\pm0.0005$ for the parameters shown in the figure. To derive a
method which can estimate $n\p\cl$ even from data for only a single
system size $L$, we use the fact that $\DP$ is $O(1)$ and scales linearly
with $n\p-n\p\cl$ to leading order near the cloud point, and hence $\ln
r\sim L^d(n\p-n\p\cl)$. This applies for $n\p<n\p\cl$, while above
$n\p\cl$ one has $\ln r=O(1)$. Thus the derivative $(\partial/\partial
n\p)\ln r$ should drop from an $O(L^d)$ plateau to $O(1)$ around
$n\p=n\p\cl$. In the second derivative $-(\partial/\partial n\p)^2\ln
r$ this drop will manifest itself as a peak. A smooth derivative can be
extracted from simulation data using histogram reweighting, and the
peak position then serves as an estimate for $n\p\cl$. This is shown in
the inset of Fig.~\ref{fig:fss}, and gives $n\p\cl\approx 0.0823$ from
the largest $L$, consistent with our earlier estimate derived from comparing
data for different $L$.

The above arguments can be formalized using the results
of~\cite{BORGS92}, which pertain to the monodisperse case but which we
have generalized to polydisperse systems~\cite{FORTHCOMING}. We find
that for large $L$ the second-derivative plot approaches a {\em universal}
master curve

\begin{equation}
-\left(\frac{\partial}{\partial\tilde{n}\p}\right)^2 \ln r = \frac{z}{(1+z)^3},
\quad
\tilde{n}\p = z+\ln z\:,
\label{eq:master}
\end{equation}
parameterized by $z$. The scaled parent density is defined here as
$\tilde{n}\p = a L^d (n\p-n\p\cl) + \ln(b L^d n\p\cl)$ with $a$ and
$b$ system-specific dimensionless scale factors. This scaling
implies that the cloud point estimate from the peak position has
finite-size corrections of order $L^{-d}\ln L$, while the peak width
and height scale as $L^{-d}$ and $L^{2d}$, respectively. Our data are
consistent with the width and height scaling and with the dominant
$L^{-d}$ dependence of the peak shifts~\cite{FORTHCOMING}.  The master
curve~(\ref{eq:master}), appropriately scaled, is overlaid onto the
largest-$L$ data in Fig.~\ref{fig:fss} (inset) and shows excellent
agreement.

Fig.~\ref{fig:coexdists} shows that the metastable liquid peak in
$p(n)$ persists until well below the cloud point $n\p\cl$. The point at
which it disappears marks the so-called mean field spinodal 
\cite{RIKVOLD94} where the liquid is unstable to small density
fluctuations. The parent density $n\p\sp$ where this occurs should
tend to an $L$-independent value as $L$ grows large \cite{RIKVOLD94},
and our data (not shown) are consistent with this. Spinodals in
monodisperse systems are conventionally characterized by the density of
the phase that becomes unstable, which is located inside the region
where phase separation occurs. Here we use instead the density $n\p\sp$
of the coexisting {\em stable} phase, which is outside this region.
This is a more meaningful representation in the polydisperse context
since only the stable (majority) phase has the parental size
distribution, while that of the metastable (minority) phase is
determined indirectly via chemical potential equality.

Equipped with a systematic method for determining cloud points, we now
consider the overall phase diagram of our system, the interparticle
potential of which was assigned the Lennard-Jones form:

\begin{equation}
u_{ij}=\epsilon_{ij}\left[\left({\sigma_{ij}}/{r_{ij}}\right)^{12}
-\left({\sigma_{ij}}/{r_{ij}}\right)^{6}\right]
\label{uij}
\end{equation}
with $\epsilon_{ij}=\sigma_i\sigma_j$,
$\sigma_{ij}=(\sigma_i+\sigma_j)/2$ and $r_{ij}=|{\bf r}_i-{\bf
r}_j|$. The potential was truncated for $r_{ij}>2.5\sigma_{ij}$ and no
tail corrections were applied. The diameters $\sigma$ are drawn from a
(parental) Schulz distribution
$f\p(\sigma)\propto\sigma^z\exp\left[(z+1)\sigma/\bar{\sigma}\right]$,
with a mean diameter $\bar{\sigma}$ which sets our unit length
scale. We chose $z=50$, corresponding to a moderate degree of
polydispersity: the standard deviation of particle sizes is
$\delta\equiv 1/\sqrt{z+1}\approx 14\%$ of the mean. The distribution
$f\p(\sigma)$ was limited to within the range
$0.5<\sigma<\sigma_c$. The upper cutoff $\sigma_c$ serves to
prevent the appearance of arbitrarily large particles in the
simulation, but would also be expected in experiment because
in the chemical synthesis of colloid particles, time or solute limits
restrict the maximum particle size \cite{KVITEK05}.

We complement the simulations with theoretical phase behaviour
calculations, following closely our study of the purely
size-polydisperse case~\cite{WILDING04}. An accurate expression for the
excess free energy of a polydisperse hard sphere fluid accounts for the
repulsive interactions. To this is added a van der Waals term which
represents the attractive part of the $u_{ij}$. It scales as
\begin{equation} 
\int
d\sigma\,d\sigma'\,\rho(\sigma)\rho(\sigma')\,(\sigma\sigma')(\sigma+\sigma')^3
\end{equation}
where the factors $\sigma\sigma'$ and $(\sigma+\sigma')^3$ arise,
respectively, from the size dependence of the interaction amplitude
$\epsilon_{ij}$ and the interaction range $\sigma_{ij}$. Multiplying out
gives an expression involving only the moment densities $\int
d\sigma\,\rho(\sigma) \sigma^{i}$ with $i=1\ldots 4$. Since the
repulsive part of the excess free energy has a similar moment
structure, the moment free energy (MFE) method~\cite{AdvChemPhys} can be used
for accurate numerical prediction of phase behaviour~\cite{WILDING04}.

\begin{figure}[h]
\includegraphics[width=8.0cm,clip=true]{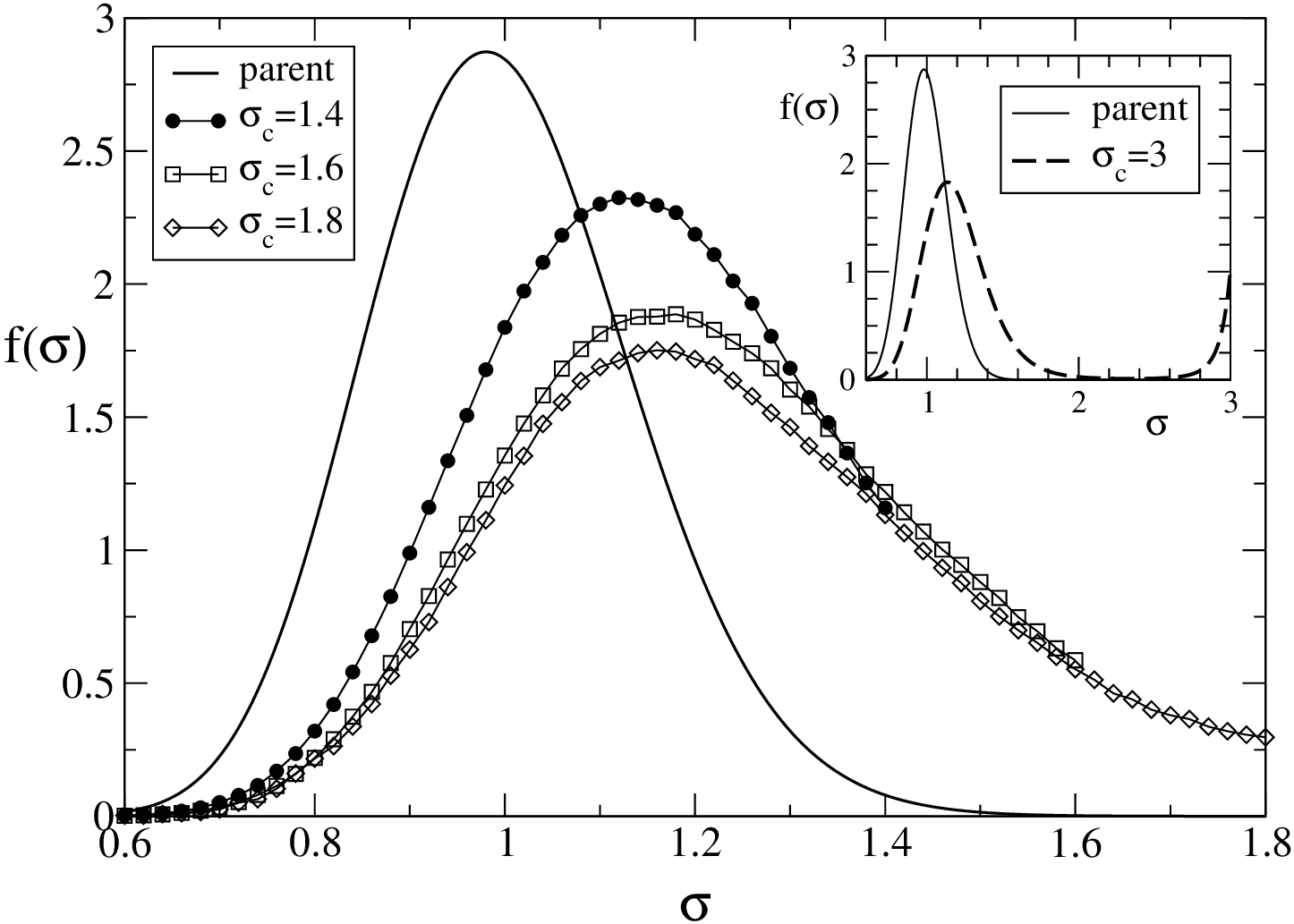}
\caption{Size distributions $f(\sigma)$ in the liquid shadow phase
distributions at $T=T_c$ for $\sigma_c=1.4, 1.6, 1.8$, together with
the parent distribution $f\p(\sigma)$. Inset: MFE theory prediction
for larger cutoff $\sigma_c=3$; note the second peak in the shadow
distribution.
}
\label{fig:shadowdist}
\end{figure}

Fig.~\ref{fig:cloudcurve} shows cloud curves for upper size cutoff
$\sigma_c=1.4$ and $1.6$ as obtained from the GCE simulations. A strong
$\sigma_c$-dependence is seen even though both values of $\sigma_c$ are
far in the tail of the parent distribution. This is attributable to
very strong fractionation effects (Fig.~\ref{fig:shadowdist}): despite
particle sizes around $\sigma_c$ being very rare in the parent, they
occur in significant concentration in the shadow liquid. Physically,
since large particles interact more strongly, their presence leads to 
a substantial free energy gain at the shorter interparticle separations
of the liquid.  

One is led to enquire whether the gas phase cloud point density would
eventually tend to a nonzero limit as $\sigma_c$ is increased. The
inset of Fig.~\ref{fig:cloudcurve} shows the simulation results and
theoretical predictions. The former exhibit a further strong decrease
of $n\p\cl$ by $\approx 25\%$ between $\sigma_c=1.6$ and $1.8$; the
latter suggest that this trend continues and that the cloud point
density tends to zero for large $\sigma_c$. Such an unusual effect has
previously been seen in theoretical studies of polydisperse hard rods
with wide length distributions ~\cite{Onsager} and is also predicted to
occur in solid-solid phase separation of polydisperse hard spheres
\cite{HS_longpaper}, though only for large $\sigma_c$ and distributions
with fatter than exponential tails. Here the decrease of $n\p\cl$ is
clear even for $\sigma_c$ of $O(1)$, i.e.\ of the same order as
$\bar{\sigma}$, and
scaling estimates suggest that cutoff effects occur for any size
distribution with tails heavier than a Gaussian~\cite{FORTHCOMING}. 

The physical origin of the decrease of $n\p\cl$ to zero is the
appearance (for large $\sigma_c$) of a second peak in the shadow phase
size distribution near $\sigma_c$ (Fig.~\ref{fig:shadowdist}, inset).
As with the hard rods, we expect this second peak to eventually
dominate as $\sigma_c$ increases so that the shadow phase consists of ever
more strongly interacting particles whose sizes are concentrated near
$\sigma_c$. We speculate that as a consequence, there exists some
cutoff for which the shadow phase liquid freezes into a
quasi-monodisperse crystal phase. Indeed our simulations provide
evidence for this scenario: for the large cutoff
$\sigma_c=2.8$ the liquid spontaneously freezes to an f.c.c.\
crystal structure \cite{FORTHCOMING}. Although we observe this only for
small $n\p$ values close to the spinodal point it is conceivable that,
for $\sigma_c$ values larger than those presently accessible to
simulations, the freezing might occur from the stable liquid phase. 

Finally,  with regard to the cloud curves as a whole
(Fig.~\ref{fig:cloudcurve}), we note that significant cutoff-dependent
shifts occur only for densities below the critical density. This is
consistent with our interpretation above: for higher densities, the
shadow phase is a gas of {\em lower} density than the parent. In this,
the concentration of large particles is {\em suppressed} and that of
small particles negligibly enhanced because of their weak interactions.
The shadow size distributions are therefore concentrated well within
the range $0.5\ldots \sigma_c$ (data not shown) so that no cutoff
dependence arises.

In summary, the task of accurately locating cloud points of
polydisperse fluids via simulation is severely complicated by
fractionation effects. We have presented a generally applicable
finite-size scaling method which addresses this problem. Application to
a model polydisperse fluid reveals the cloud curve to be highly
sensitive to the presence of very rare large particles. Such effects
imply that in experiments on polydisperse fluids (see e.g.~\cite{ERNE05})
it may be important to monitor and control carefully the tails of the
size (or charge, etc) distribution. Otherwise undetected differences
could lead to large sample-to-sample fluctuations in the observed phase
behaviour.


\end{document}